# Influence Maximization based on Simplicial Contagion Models in Hypergraphs

**Renquan Zhang,[1] Ting Wei,[1] Yifan Sun,[2] and Sen Pei[3,*]**

1) *School of Mathematical Sciences, Dalian University of Technology*
2) *Center for Applied Statistics and School of Statistics, Remin University of China, Beijing, China*
3) *Department of Environmental Health Sciences, Mailman School of Public Health, Columbia University*

(*Electronic mail: sp3449@cumc.columbia.edu.)

(Dated: 21 November 2023)

In recent years, the exploration of node centrality has received significant attention and extensive investigation, primarily fuelled by its applications in diverse domains such as product recommendations, opinion propagation, disease spread, and other scenarios requiring the maximization of node influence. Despite various perspectives emphasizing the indispensability of higher-order networks, research specifically delving into node centrality within the realm of hypergraphs has been relatively constrained. This paper focuses on the problem of influence maximization on the Simplicial Contagion Model (SCM), using the susceptible-infected-recovered (SIR) model as an example. To find practical solutions to this optimization problem, we have developed a theoretical framework based on message passing process and conducted stability analysis of equilibrium solutions for the self-consistent equations. Furthermore, we introduce a metric called collective influence and propose an adaptive algorithm, known as the Collective Influence Adaptive (CIA), to identify influential propagators in the spreading process. Notably, our algorithm distinguishes itself by prioritizing collective influence over individual influence, resulting in demonstrably superior performance, a characteristic substantiated by a comprehensive array of experiments.

This paper presents innovative contributions across multiple dimensions. Firstly, we extend the Message Passing equation to accommodate Simplicial Contagion Models on hypergraphs. Secondly, the introduction of the Collective Influence concept quantifies the collective impact within hypergraphs. Thirdly, experimental observations reveal that nodes with higher Collective Influence (CI) values exhibit a pronounced clustering tendency. Building upon this empirical insight, our CIA algorithm strategically leverages this trend to mitigate excessive node clustering within optimal sets. Extensive simulations conducted on both synthetic and real-world hypergraphs substantiate the superior performance of the CIA algorithm when compared to classical algorithms. Furthermore, our investigation reveals a noteworthy phenomenon in this propagation model, highlighting the significant role played by low-order interactions (1-simplex). These findings contribute valuable insights into the influence of hypergraph characteristics on the efficacy of the CIA algorithm.

## I. INTRODUCTION

In complex networks, the propagation process can describe various real phenomena, including the spread of diseases [1-3], ideas [4-9], or products among people [10,11]. Due to the structural heterogeneity of networks, a small number of nodes play a disproportionate role in shaping the outcome of propagation dynamics. Identifying these key nodes is a critical issue in network science. The optimal influence problem was initially introduced in the context of viral marketing, aiming to increase global influence with a small number of targeted interventions. In recent years, the Influence Maximization (IM) problem on simple graphs has been widely studied [12-15].

However, in many real scenarios, interactions often involve more than two individuals. For example, in social systems [16], neuroscience [17,18], ecology [19], and biology [20], many connections and relationships do not take place between pairs of nodes, but rather are collective actions at the level of groups

Influence Maximization in Hypergraphs

of nodes. A scientific paper may have more than two co-authors, and multiple users may form groups on social platforms for information sharing. Such relationships can be represented using hypergraphs, where hyperedges describe the interactions between more than two nodes [21]. The IM problem on hypergraphs (HIM) remains largely unexplored, with only a few studies focusing on this area. In 2019, Zhu et al. first investigated information diffusion in social networks abstracted as directed hypergraphs, proving that the problem under the independent cascade model is NP-hard and proposing an approximation algorithm [22]. Antelmi et al. developed three greedy heuristic methods to address this problem [23]. Although the aforementioned techniques are applicable to certain specific situations, the HIM problem still faces the challenge of balancing effectiveness and efficiency. Algorithms that obtain optimal solutions often require a large amount of time, making them difficult to apply to large-scale hypergraphs. Heuristic algorithms, on the other hand, do not optimize the global influence function and cannot guarantee their performance. Hence, there is a need to devise efficient methodologies that can closely approximate the optimal solution for the diffusion of seed nodes.

In this study, we focus on the influence maximization problem in the Simplicial Contagion Model (SCM) proposed by Iacopini et al. [24]. To address this optimization problem, we first develop a mathematical framework that represents the system as a message passing process, in which the focal nodes are assumed to be "virtually" removed from the hypergraph, and then perform stability analysis of the equilibrium solutions. The critical state is therefore governed by the largest eigenvalue of the weighted non-backtracking (WNB) matrix. Integral to our methodology is the introduction of a novel, theory-based metric termed "collective influence" designed to quantitatively measure the impact of each node within the SCM. Subsequently, we propose an adaptive algorithm based on collective influence (CIA), validate the proposed method in both synthetic and real-world networks, and demonstrate that it outperforms commonly used approaches.

## II. PRELIMINARY DEFINITION

This section extends the definitions from simple graphs to hypergraphs, introducing the SCM, and providing the definition of the HIM problem.

**A.** Definition of a hypergraph

Hypergraphs are represented as $H(V, E)$, where $V = \{v_1, v_2 \ldots v_N\}$ denotes the set of nodes and $E = \{e_1, e_2 \ldots e_M\}$ represents the set of hyperedges. Each hyperedge $e_i = \{v_{i_1}, v_{i_2} \ldots v_{i_k}\}$ is a subset of the node set, representing group interactions among nodes. The incidence matrix of hypergraph $H = (V, E)$ is an $N \times M$ matrix $I = \{I_{i\alpha}\}$. The entry $I_{i\alpha}$ in row $i$ and column $\alpha$ is 1 if node $i$ and edge $\alpha$ are incident, and 0 otherwise [25]. Therefore, the adjacency matrix of the hypergraph can be defined as:

$$A = II^T - D, \qquad (1)$$

where $D$ is a diagonal matrix with diagonal entries representing the number of hyperedges to which each node $i$ belongs.

To represent the connectivity between node $v_i$ and node $v_j$, the adjacency matrix $A$ can be defined as a binary matrix $\tilde{A}$, where the element $A_{ij} = 1$ if node $v_i$ and node $v_j$ share at least one hyperedge, and 0 otherwise. Specifically, it can be defined as follows:

$$\tilde{A}_{ij} = \begin{cases} 1, & A_{ij} \geq 0 \\ 0, & A_{ij} = 0 \end{cases} \qquad (2)$$

An example of a hypergraph is given in Figure 1, which contains 5 nodes and 4 hyperedges. The incidence matrix $I$, adjacency matrix $A$, and binary adjacency matrix $\tilde{A}$ are also provided accordingly. Similarly, we can define the adjacency tensor $\tilde{B}$ to represent the existence of a 2-simplex in a hypergraph, as well as the weighted adjacency tensor $B$. Given the adjacency matrix and incidence matrix of a hypergraph, the degree and hyperdegree of a node are defined as follows.

The hyperdegree of a node $v_i$ indicates the number of hyperedges to which node $v_i$ belongs, which is formally defined as:

$$d_H(i) = \sum_{\alpha=1}^{M} I_{i\alpha} \qquad (3)$$

The degree of node $v_i$ is defined as the number of

# Influence Maximization in Hypergraphs

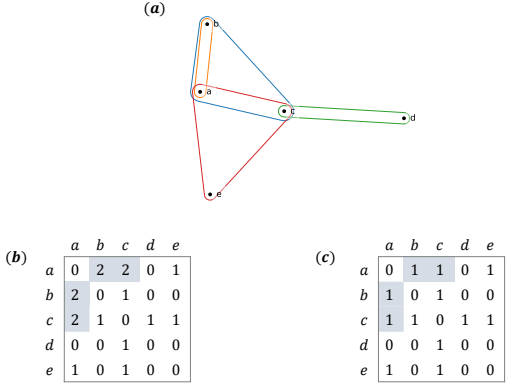

**Figure 1.** An illustration example of $(a)$ a hypergraph; $(b)$ the adjacency matrix $A$ of $(a)$; $(c)$ the binarized adjacency matrix $\tilde{A}$ of $(a)$.

its neighbors:

$$d_N(i) = \sum_{j=1}^{N} \tilde{A}_{ij} \quad (4)$$

The size of a hyperedge $e_\alpha$ is defined as:

$$d_E(\alpha) = \sum_{i=1}^{N} I_{i\alpha} \quad (5)$$

**B. The contagion model**

A simplicial complex is a special type of hypergraph that includes all non-empty subsets as hyperedges. In the SCM, the contagion process is considered to occur through either links or collective interactions, with different propagation probabilities. The infection probability $B = \{\beta_1, \beta_2 \ldots \beta_R\}$ for each element in a simplicial complex represents the probability that a susceptible node $i$ will be infected through each $R$-dimensional subface of the complex in a unit time, where $R \in [1, N-1]$. Thanks to the simplicial complex requirements that all subsimplices of a simplex are included, contagion processes in a $R$-simplex among which $D < R$ nodes are infectious are also automatically considered. For simplicity, we only consider dimensions up to $R = 2$. Figure 2(a) serves as an illustrative diagram.

We use the susceptible–infectious–recovered (SIR) model to simulate the spreading process on a hypergraph consisting of $N$ nodes and $M$ undirected hyperedges. We associate three binary state variables $S_i(t)$, $I_i(t)$, and $R_i(t)$ with the $N$ nodes of the hypergraph $H$, representing the susceptible, infected, and recovered states of node $i$ at time $t$, respectively. We denote the probability of an infected node infecting its susceptible neighbor in a 1-simplex as $\beta_1$, and the probability of the last node being infected when two nodes are infected in a 2-simplex as $\beta_2$, and define $\gamma$ as the recovery period (without loss of generality, $\gamma = 1$).

In each time step $t$, a susceptible node $i$ ($S_i(t) = 1$) can be infected by its infected neighbor $j$ ($I_j(t) = 1$) through a link (1-simplex) $(i,j)$ with infection probability $\beta_1$. Similarly, a susceptible node $i$ can also be infected by its infected neighbors $j$ and $k$ ($I_j(t) = 1, I_k(t) = 1$) through a triangle (2-simplex) $(i,j,k)$ with infection probability $\beta_2$. Meanwhile, an infected node will recover after $\gamma$ steps and become immune to further infection. We show an example of SIR diffusion process in the SCM in Figure 2(a). We generated 10 scale-free hypergraphs with 1000 nodes each, randomly selected seed, and conducted 100 independent experiments. The results are shown in Figure 2(b), where the average fraction of infected nodes, obtained from the simulations, is plotted against the rescaled infectivity $\lambda = \frac{\beta \langle k \rangle}{\mu}$ for a ($R = 2$) SCM with $\lambda_2 = 0.8$ (yellow circles), $\lambda_2 = 2.5$ (blue squares), and $\lambda_2 = 0$ (red triangles). The increase from $\lambda_2 = 0$ to $\lambda_2 = 0.8$ leads to a slight increment in the infection rate, a phenomenon that will be elucidated in Section III.B. It has been shown that the inclusion of higher-order interactions leads to the emergence of new phenomena. This alters the nature of the transition at the epidemic threshold from continuous to discontinuous and results in the appearance of a bistable region in the parameter space, where both healthy and endemic asymptotic states coexist. Figure 2(c) illustrates the distribution of absorbing and endemic states for $\lambda_1 = 1$ and $\lambda_2 = 2.5$. It shows that 34.01% of the population is in the absorbing states, while 65.99% is in the endemic states.

The evolution of these probabilities can be described as follows:

Influence Maximization in Hypergraphs

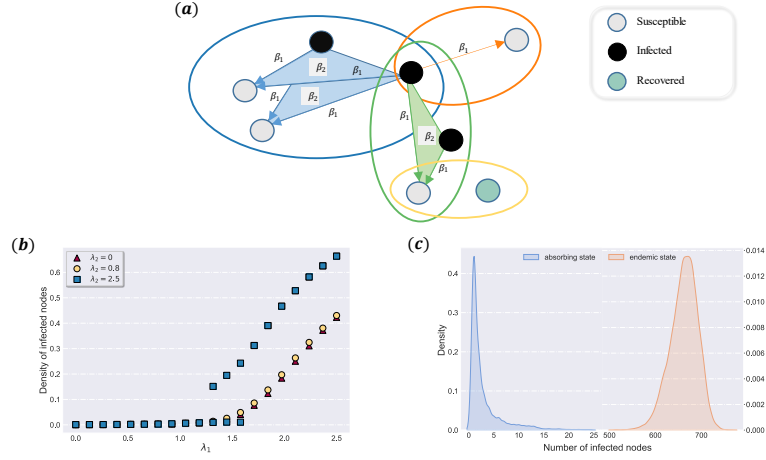

**Figure 2.** ($a$) Schematic illustration of SIR diffusion in the SCM. Infected nodes (filled black circles) infect a healthy node (white circle) via 1-simplex and 2-simplex with rates $\beta_1$ and $\beta_2$, respectively. The infected nodes will transition to an immune state (green circles) after $\gamma$ steps. ($b$) The experiment was conducted on 10 SF hypergraphs consisting of 1000 nodes. The numerical results of the standard SIR model without considering higher-order effects ($\lambda_2 = 0$) are represented by red triangles. The transition from $\lambda_2 = 0$ to $\lambda_2 = 0.8$ represents a continuous phase transition. For $\lambda_2 = 2.5$ we observe a discontinuous phase transition with the formation of a bistable region where absorbing and endemic states co-exist. ($c$) In the parameter region of bistable ($\lambda_1 = 1, \lambda_2 = 2.5$), the distribution of the two states.

$$\frac{dS_i(t)}{dt} = -S_i(t)\left[1 - \prod_j \left(1 - \beta_1 \tilde{A}_{ij} I_j(t)\right)^{A_{ij}} \right.$$
$$\left. \times \prod_{k,l} \left(1 - \beta_2 \tilde{B}_{ikl} I_k(t) I_l(t)\right)^{B_{ikl}} \right] \quad (6)$$

$$\frac{dI_i(t)}{dt} = S_i(t)\left[1 - \prod_j \left(1 - \beta_1 \tilde{A}_{ij} I_j(t)\right)^{A_{ij}} \right.$$
$$\left. \times \prod_{k,l} \left(1 - \beta_2 \tilde{B}_{ikl} I_k(t) I_l(t)\right)^{B_{ikl}} \right] - \frac{I_i(t)}{\gamma} \quad (7)$$

$$\frac{dR_i(t)}{dt} = \frac{I_i(t)}{\gamma} \quad (8)$$

**C.** Problem statement

This study mainly addresses the problem of hypergraph influence maximization (HIM), aiming to identify the optimal set $S = \{v_1, v_2 ... v_k\}$ of $k$ influential spreaders in a hypergraph under a specific diffusion mechanism, in order to maximize the expected influence $\sigma(S)$. The HIM problem can be seen as a generalization of the IM problem on hypergraphs and is also NP-hard. This means that it cannot be solved in polynomial time. Indeed, finding the optimal set of influencers is a many-body problem in which the topological interactions between them play a crucial role. Existing research either converts the hypergraph into a bipartite graph or designs heuristic algorithms to solve the influence maximization problem on hypergraphs. However, these heuristic methods do not consider the global influence function. We seek the optimal node set $S$ by quantifying the expected influence of each node through collective influence.

**III. METHOD**

**A.** Message passing equations

In simple graphs, the message passing equations enhance node independence and simplify the definition of information propagation by removing a node from the network, thereby greatly minimizing the occurrence of loops. We extend this concept to the SCM by generalizing the message passing equations.

Influence Maximization in Hypergraphs

Compared to the master equation defined using the adjacency matrix, the message passing process better captures the dynamics of the SIR model. Specifically, in the SIR model, backtracking infection ($i \to j \to i$) is not allowed as the propagation is irreversible. Previous research has also shown the superiority of the message passing process in analyzing dynamical models in complex networks [26-34].

To study the impact of a node $j$ on its neighbor node $i$, we investigate the probability of node $i$ being infected if node $j$ is assumed to be absent from the hypergraph. For a link from $i$ to $j$ ($i \to j$), suppose node $j$ is "virtually" removed from the hypergraph (i.e., creating a "cavity" at node $j$) and calculate the probability of node $i$ being infected in the absence of node $j$ at time $t$, which is represented as $S_{i \to j}(t)$. We apply the same procedure for I and R. The message passing process can be described by

$$S_{i \to j}(t+1) = S_{i \to j}(t) \left[ \prod_{k \backslash j} \left(1 - \beta_1 \tilde{A}_{ik} I_{k \to i}(t)\right)^{A_{ik}} \right.$$
$$\left. \times \prod_{m,l \backslash j} \left(1 - \beta_2 \tilde{B}_{iml} I_{m \to i}(t) I_{l \to i}(t)\right)^{B_{iml}} \right] \quad (9)$$

$$I_{i \to j}(t+1) = S_{i \to j}(t) \left[ 1 - \prod_{k \backslash j} \left(1 - \beta_1 \tilde{A}_{ik} I_{k \to i}(t)\right)^{A_{ik}} \right.$$
$$\left. \times \prod_{m,l \backslash j} \left(1 - \beta_2 \tilde{B}_{iml} I_{m \to i}(t) I_{l \to i}(t)\right)^{B_{iml}} \right]$$
$$+ I_{i \to j}(t) \left(1 - \frac{1}{\gamma}\right) \quad (10)$$

$$R_{i \to j}(t+1) = R_{i \to j}(t) + \frac{I_{i \to j}(t)}{\gamma} \quad (11)$$

Here $k \backslash j$ means $k$ runs over all nodes except $j$. Denote $\lim_{t \to \infty} S_{i \to j}(t) = S_{i \to j}$, $\lim_{t \to \infty} I_{i \to j}(t) = I_{i \to j}$, $\lim_{t \to \infty} R_{i \to j}(t) = R_{i \to j}$. The steady state of the nonlinear dynamical system can be obtained by solving the following self-consistent equations:

$$S_{i \to j} = S_{i \to j} \left[ \prod_{k \backslash j} \left(1 - \beta_1 \tilde{A}_{ik} I_{k \to i}\right)^{A_{ik}} \right.$$
$$\left. \times \prod_{m,l \backslash j} \left(1 - \beta_2 \tilde{B}_{iml} I_{m \to i} I_{l \to i}\right)^{B_{iml}} \right] \quad (12)$$

$$I_{i \to j} = \gamma S_{i \to j} \left[ 1 - \prod_{k \backslash j} \left(1 - \beta_1 \tilde{A}_{ik} I_{k \to i}\right)^{A_{ik}} \right.$$
$$\left. \times \prod_{m,l \backslash j} \left(1 - \beta_2 \tilde{B}_{iml} I_{m \to i} I_{l \to i}\right)^{B_{iml}} \right] \quad (13)$$

**B. Stability analysis**

The above self-consistent equation has a trivial solution $\left(S^*_{i \to j}, I^*_{i \to j}\right)^T = (1,0)^T$, which is that all nodes in the hypergraph are in the susceptible state. The stability of this trivial solution is determined by the maximum eigenvalue of the Jacobian matrix, **J**, at this solution. For a given hypergraph, we take the partial derivative of Eq. (12) and for directed links $k \to l$ and $i \to j$, we have:

$$\frac{\partial S_{i \to j}}{\partial S_{k \to l}} = \begin{cases} 1 & i = k, j = l \\ 0 & \text{otherwise} \end{cases} \quad (14)$$

$$\frac{\partial S_{i \to j}}{\partial I_{k \to l}} = \begin{cases} -\beta_1 A_{ik} & l = i, k \neq j \\ 0 & \text{otherwise} \end{cases} \quad (15)$$

The same analysis on Eq. (13) yields:

$$\frac{\partial I_{i \to j}}{\partial S_{k \to l}} = 0 \quad (16)$$

$$\frac{\partial I_{i \to j}}{\partial I_{k \to l}} = \begin{cases} \beta_1 \gamma A_{ik} & l = i, k \neq j \\ 0 & \text{otherwise} \end{cases} \quad (17)$$

So the Jacobian matrix at the solution $(1,0)^T$ is given by:

$$\mathbf{J}|_{(1,0)} = \begin{pmatrix} \mathbf{I} & \mathbf{B} \\ \mathbf{0} & \mathbf{C} \end{pmatrix} \quad (18)$$

Here **I** is the identity matrix, $\mathbf{B} = \left\{\frac{\partial S_{i \to j}}{\partial I_{k \to l}}\right\}_{2M \times 2M}$ and $\mathbf{C} = \left\{\frac{\partial I_{i \to j}}{\partial I_{k \to l}}\right\}_{2M \times 2M}$, where $M$ is the number of links. The matrix **C** is a generalization of the non-backtracking (NB) matrix of networks **N**, which is a weighted non-backtracking (WNB) matrix, where

$$\mathbf{N}_{k \to l, i \to j} = \begin{cases} 1 & l = i, k \neq j \\ 0 & \text{otherwise} \end{cases} \quad (19)$$

The stability of $I^*_{i \to j}$ is determined by the largest eigenvalue of the matrix **C**, denoted by $\lambda_C$. The trivial solution is stable if $\lambda_C < 1$ and unstable if $\lambda_C > 1$. The HIM problem can be mapped to finding the node

Influence Maximization in Hypergraphs

set that maximally reduces the largest eigenvalue of the matrix $\mathbf{C}$. Moreover, from the Jacobian matrix $\mathbf{J}$, we know that low-order interactions (1-simplex) play a major role in this propagation model.

Denote the leading eigenvector of $\mathbf{C}$ as $\mathbf{v}$ such that $\mathbf{Cv} = \lambda\mathbf{v}$, where $\lambda$ is the largest eigenvalue of $\mathbf{C}$. The eigenvalue $\lambda$ determines the growth rate of any non-zero vector $\mathbf{v}$ of the matrix $\mathbf{C}$ after $l$ iterations. The high dimensionality of the matrix $\mathbf{C}$ makes it difficult to calculate the leading eigenvalue of large-scale networks. Power iteration is an effective method for computing the principal eigenvalue and eigenvector. Following the method in Ref.[35], we derive that the $CI_l(i)$ of a node $i$ in the network when $l = 1$ is given by:

$$CI_1(i) = (\beta_1\gamma)^2 \sum_{j\in Ball(i,1)} A_{ij}z_i^j(d_N(j) - 1) \quad (20)$$

where $z_i^j = \sum_{k\backslash j} A_{ik}$.

**C. Adaptive algorithm based on collective influence**

The collective influence of node $i$ encompasses not only its own importance but also considers the importance of its neighbors $j$ ($j \in Ball(i,1)$), where nodes with elevated CI values tend to be associated with neighbors exhibiting correspondingly heightened CI values. To empirically validate this hypothesis, we conducted experiments on both synthetic and real-world networks. As illustrated in Figure 3, the depicted probability denotes that the neighbors of the top $n\%$ nodes concurrently occupy the top $n\%$ positions. This phenomenon is notably pronounced across a spectrum of real-world networks. Moreover, the status of these nodes is usually symmetric, and selecting these neighbors as seeds can result in multiple infections of the same nodes. Consequently, the designation of a solitary node from this cohort suffices as an effective seed selection strategy.

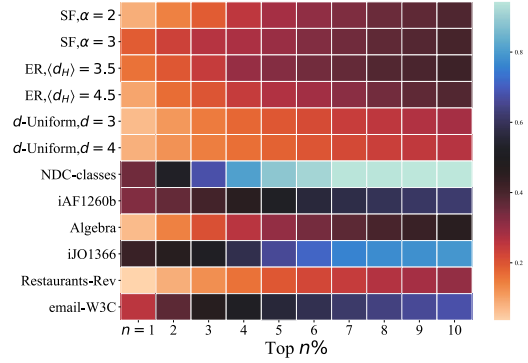

**Figure 3.** The probability that the neighbors of the top $n\%$ of nodes in ER hypergraph, SF hypergraph, $d$-Uniform hypergraph and six real-world hypergraphs are also in the top $n\%$.

Based on this assumption, we propose an adaptive algorithm based on collective influence, i.e., CIA, to solve the HIM problem.

| Algorithm 1 Collective Influence Adaptive |
|---|
| **Input:** Size of seed nodes $k$ |
|        Hypergraph $H(V, E)$ |
| **Output:** Seed node set $S$ |
| **Initialization:** $S = \emptyset, L \leftarrow V$ |
| Compute $CI_i$ for $1 \leq i \leq N$ according to Eq.(20) |
| **while** $\|S\| < k$ **do** |
|     $v_j(v_j \in L\backslash S) \leftarrow max\{CI_j\}$ |
|     $N(S) \leftarrow$ Neighbors of nodes in Seed node set $S$ |
|     **if** ($v_j$ in $N(S)$) **then** |
|         $L \backslash v_j$ |
|     **else** |
|         $S \leftarrow S \cup \{v_j\}$ |
|         $L \backslash v_j$ |
| **end** |

## IV. NUMERICAL VALIDATION

To test the performance of the adaptive algorithm based on collective influence, we conducted numerical simulations on synthetic and real-world hypergraphs respectively. We selected spreaders using different methods and conducted 100 independent experiments on the SIR model. The infection size $\sigma(S)$ was taken as the average of 100 simulations to ensure reliability of the results. We compared the CIA with several competing methods, including (1) degree-based ranking, (2) hyperdegree-based ranking, (3) a naive extension of CI on simple graphs, (4) HADP algorithm, and (5) HSDP algorithm [36]. We also tested random algorithm as a baseline. More details of these competing methods are provided in Appendix A. In addition to altering the network structure, we conducted the following experiments: (a) varying $\beta_1$, (b) varying $\beta_2$, (c) altering the network scale $N$, and (d) adjusting the number of initial seeds $k$, to validate the effectiveness of our approach under different infection rates and initial conditions, as well as to ascertain its applicability to large-scale networks. Illustrations are provided in Figure 4. Our results show that the CIA algorithm has significant advantages over other algorithms.

**A.** Scale-Free hypergraphs

To check the robustness of our method over the change of the degree heterogeneity, the performance of our methods on Scale-Free (SF) hypergraphs is evaluated. We used the Chung- Lu model to generate hypergraphs with given hyperdegree and hyperedge size distributions, where the hyperdegree sequence was generated by the distribution $p(d_H) \sim (d_H)^{-\alpha}$ and the hyperedge size sequence was generated by the distribution $p(d_E) \sim (d_E)^{-\alpha}$, with the exponent $\alpha$ being a tunable parameter. In this paper, we set the exponent $\alpha = 2$ and 3.

We used synthetic hypergraph with $\alpha = 2$ or 3 and $N = 1000, 5000$, and $10{,}000$. To ensure the connectivity of the hypergraphs, all simulations were applied only on the giant connected component (GCC). According to Ref.[24], we can define the 1-simplex density $\langle k_1 \rangle$ and the 2-simplex density $\langle k_2 \rangle$, and adjust the parameters $\beta_1$ and $\beta_2$ by the factors $\lambda_1 = \frac{\beta_1 \langle k_1 \rangle}{\mu}$ and $\lambda_2 = \frac{\beta_2 \langle k_2 \rangle}{\mu}$, controlling $\lambda_1 \in [0.6, 1.2]$ and $\lambda_2 \in \{1,3\}$. We did not consider higher values of $\beta_1$, as even random seed selection would lead to similar infection sizes at such high values.

We first conducted tests with a small propagation probability of high-order interactions ($\beta_2 = 0.2$), fixing the seed number $k$ to 3% of the size of the GCC, and varying $\beta_1$. In Figure 5, we compared various algorithms by evaluating the ratio of the number of infected nodes to the number of nodes in the GCC, i.e., $\sigma(S)/|V_{GCC}|$. A higher ratio indicates better performance of the method. Overall, the CIA algorithm consistently outperformed other competing methods, with HADP performing second best and the random algorithm performing the worst.

As $\beta_1$ increases, the gap between CIA algorithm and other algorithms first increases and then decreases, reaching its peak around $\beta_1 = 0.25$. On the network with $\alpha = 2$, the CIA algorithm can lead the random algorithm by up to 9.95%, 8.10%, and 8.76%, respectively, and lead the second-best HADP algorithm by up to 2.27%, 1.86%, and 2.25%, respectively. When $\alpha = 3$, the network has stronger heterogeneity, and the CIA algorithm can lead the random algorithm by up to 7.04%, 7.96%, and 7.92%, respectively, and lead the second-best HADP algorithm by up to 2.84%, 1.98%, and 1.90%, respectively.

As the network size increases, the fluctuations of all methods become smaller. In these experiments, the CIA algorithm has the largest median and the smallest fluctuation, while the random algorithm has the largest fluctuation and the smallest median. We show in Figure 6 the distribution of the proportion of 100 infections when $\beta_1 = 0.25, \beta_2 = 0.2$, and the results are similar for other cases.

In Section III.B, we have shown that high-order interactions (2-simplex) do not play a major role, and we have also conducted experiments with $\beta_2 = 0.5$. When $\alpha = 2$, as $\beta_1$ increases, CIA outperforms random algorithm by up to 9.97%, 8.32%, and 8.60%, and outperforms the second-best method HADP by up to 2.70%, 1.43%, and 1.95%, respectively. When $\alpha = 3$, CIA outperforms random algorithm by up to 7.22%, 7.79%, and 8.21%, and outperforms HADP by up to 2.62%, 2.02%, and 2.09%, respectively. Compared to the case with $\beta_2 = 0.2$, CIA still has an advantage, and the advantage is approximately equal. The results are shown in Figure B1. Similar distributions for $\beta_1 = 0.25, \beta_2 = 0.5$ can be found in Figure B2.

Next, we consider the performance of each method with an increased number of seeds. For each network, we consider the number of seeds from 10 to around 10% of the size of GCC. As shown in Figure 7, with $\lambda_1 = 1$ and $\lambda_2 = 1$ fixed, as the number of seeds increases, the performance of CIA algorithm is significantly better than other algorithms, and its lead over other methods also gradually increases. HADP performs as the second-best method, while the random

Influence Maximization in Hypergraphs

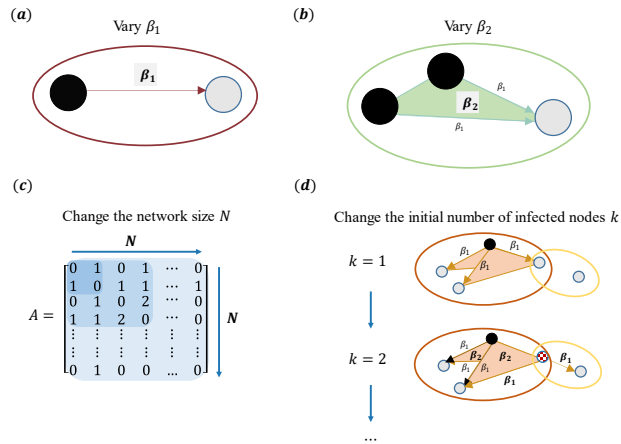

**Figure 4.** Visual illustration of changed configurations in evaluation of experiment.

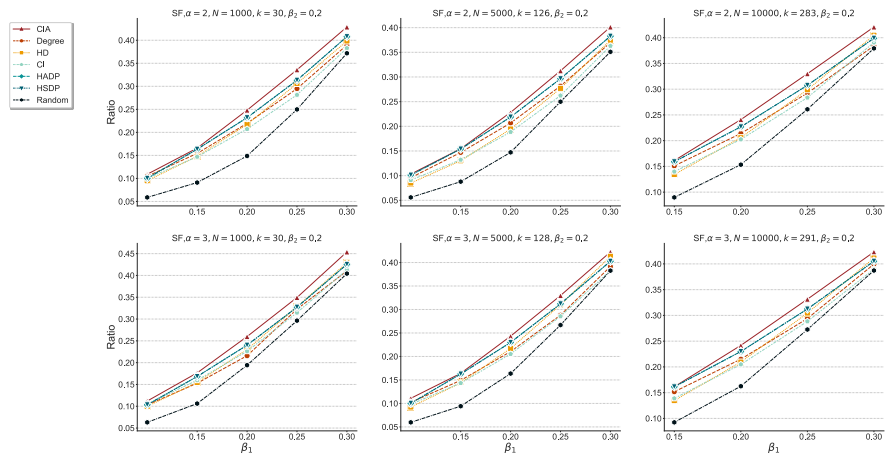

**Figure 5.** The fraction of influenced nodes with $\beta_2 = 0.2$ in SF hypergraphs of size $N = 1000, 5000$ or $10000$ and $\alpha = 2$ or $3$. We use different methods (including CIA, degree, HD, CI, HADP, HSDP and Random) to select seeds of 3% of the size of GCC. These results are the average of 100 independent experiments.

algorithm performs the worst. We also conducted experiments with $\lambda_1 \in (0.8, 1.6)$, and the experimental results are similar.

**B.** Erdös-Rényi hypergraph

We tested the proposed method on homogeneous Erdös–Rényi (ER) random networks. We generated a hypergraph with $N$ nodes, $M$ hyperedges, and a probability $p$ of connecting a node to a hyperedge in a bipartite graph. The resulting hypernetwork has an average hyperdegree of $\langle d_H \rangle = Mp$ and an average hyperedge size of $\langle d_E \rangle = Np$. We used generated networks with $N = 1000, 5000$ or $10000$ and $\langle d_N \rangle = 3.5$ or $4.5$ as shown in the Figure 8. We controlled $\beta_1 \in [0.15, 0.3]$. The performance of CIA was consistently the best, followed by HADP, while the random algorithm performed the worst.

On the hypergraph with $\langle d_N \rangle = 3.5$, CIA

Influence Maximization in Hypergraphs

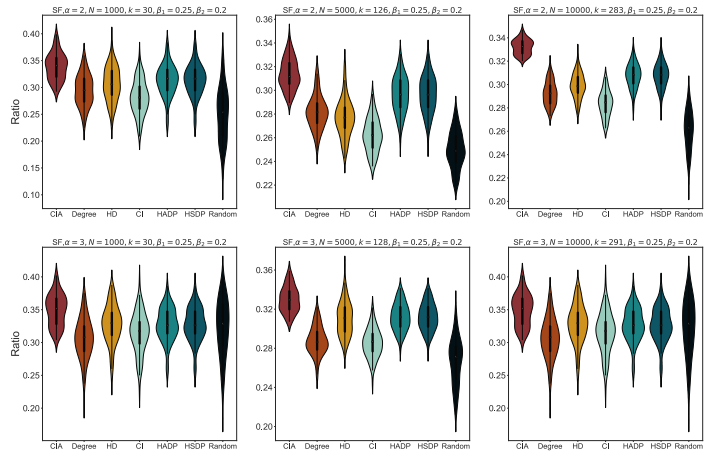

**Figure 6.** The distribution of the fraction of influenced nodes with $\beta_1 = 0.25$ and $\beta_2 = 0.2$ in SF networks of size $N = 1000, 5000$ or $10000$ and $\alpha = 2$ or $3$. We use different methods (including CIA, degree, HD, CI, HADP, HSDP and Random) to select seeds of 3% of the size of the GCC. These results are based on 100 independent experiments.

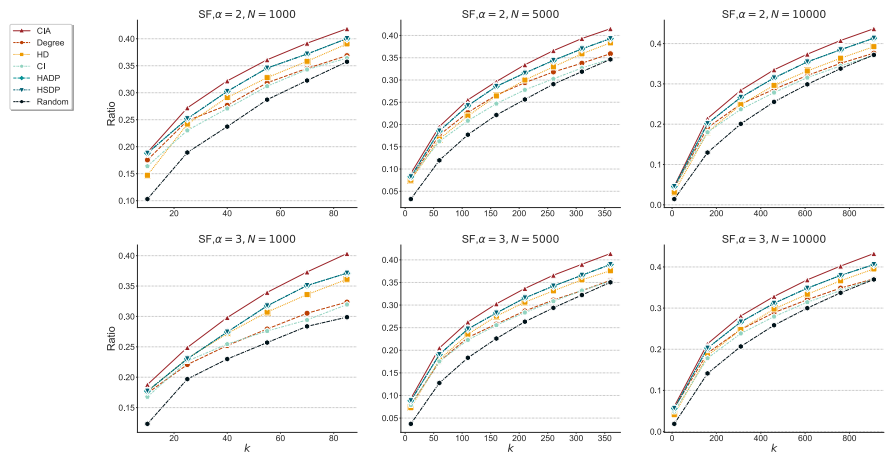

**Figure 7.** The proportion of infected nodes with increasing seed numbers in SF networks of size $N = 1000, 5000$ or $10000$ with $\alpha = 2$ or $3$, and fixed $\lambda_1 = 1$ and $\lambda_2 = 1$. These results are the average of 100 independent experiments.

outperforms the random algorithm by up to 8.41%, 8.62%, and 8.41%, and outperforms the second-best method HADP by up to 1.72%, 1.22%, and 1.18%, respectively. When $\langle d_N \rangle = 4.5$, the hypergraph is denser, and CIA outperforms the random algorithm by up to 7.03%, 6.79%, and 7.09%, and outperforms HADP by up to 0.67%, 1.64%, and 0.79%, respectively. The results show that as the average degree increases, the differences between various algorithms decrease, and our algorithm performs better in sparser networks.

Similar to the SF hypergraph, the median of the CIA algorithm is the highest and has the smallest fluctuations, while the random algorithm has the largest fluctuations and the smallest median, and its distribution is the most dispersed. We show the distribution of the proportion of infected nodes in 100 experiments with $\beta_1 = 0.3$ and $\beta_2 = 0.2$ in Figure 9, and other results are similar. Similar experiments were also conducted at $\beta_2 = 0.5$, and the results are shown

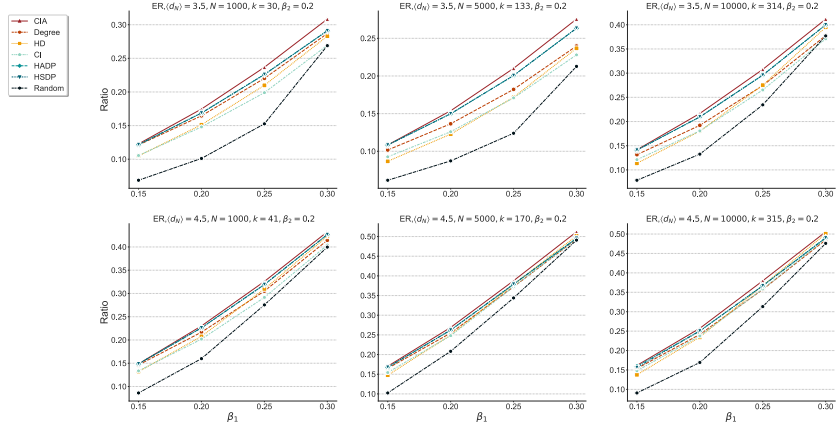

**Figure 8.** The fraction of influenced nodes with $\beta_2 = 0.2$ in ER hypergraphs of size $N = 1000, 5000$ or $10000$ and $\langle d_N \rangle =$ 3.5 or 4.5. We use different methods (including CIA, degree, HD, CI, HADP, HSDP and Random) to select seeds of 3% of the size of GCC. These results are the average of 100 independent experiments.

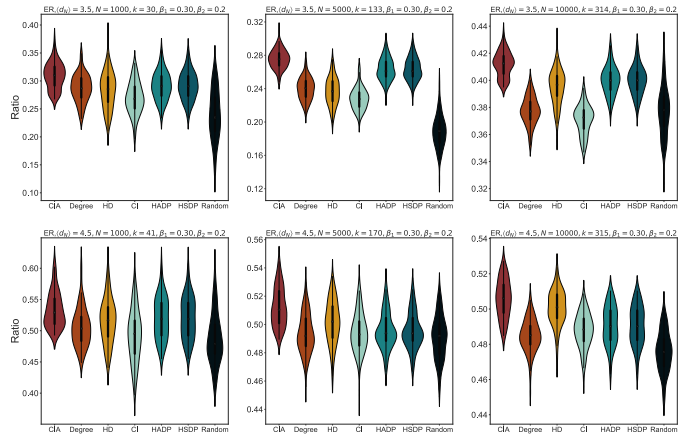

**Figure 9.** The distribution of the fraction of influenced nodes with $\beta_1 = 0.30$ and $\beta_2 = 0.2$ in ER networks of size $N = 1000, 5000$ or $10000$ and $\langle d_N \rangle = 3.5$ or 4.5. We use different methods (including CIA, degree, HD, CI, HADP, HSDP and Random) to select seeds of 3% of the size of the GCC. These results are based on 100 independent experiments.

in Figure B3 and Figure B4.

As shown in Figure 10, after increasing the number of seeds, the performance of CIA remains the best, and the gap between CIA and the random algorithm first increases and then decreases, while the advantage of CIA over other algorithms gradually increases. This is because when the number of seeds exceeds a certain proportion, even randomly selecting seeds can lead to an infection of the same size. Moreover, similar to the experiment of increasing $\beta_1$, the advantage of CIA is more significant in hypergraphs with $\langle d_N \rangle = 3.5$.

**C.** Real-world hypergraph

Finally, we evaluated the CIA algorithm on six real-world hypergraphs to demonstrate its effectiveness. Their topological properties are shown in Table 1. Similar to the generated networks, we conducted 100 independent experiments on the SIR model starting from the selected seed nodes, with $\lambda_1 = 1.1$ and $\lambda_2 = 1$ fixed. We also increased the number of seed nodes and the proportion of infected nodes

Influence Maximization in Hypergraphs

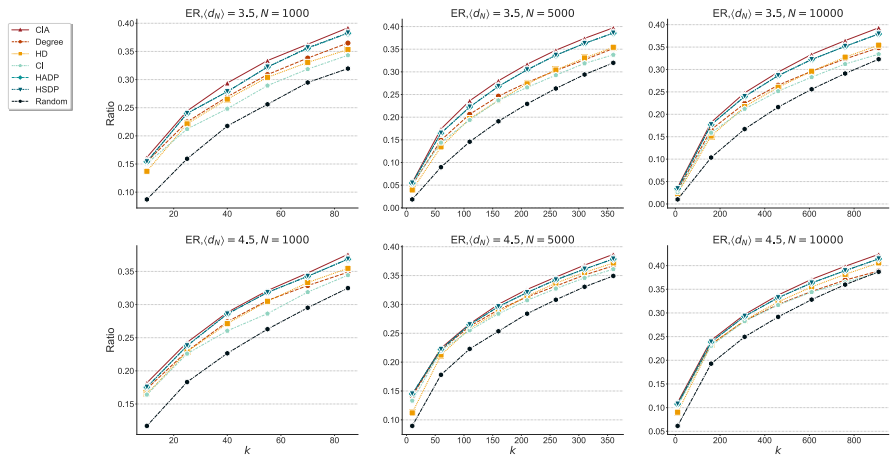

**Figure 10.** The proportion of infected nodes with increasing seed numbers in ER networks of size $N = 1000, 5000$ or $10000$ with $\langle d_N \rangle = 3.5$ or $4.5$, and fixed $\lambda_1 = 1$ and $\lambda_2 = 1$. These results are the average of 100 independent experiments.

**Table 1.** Statistics of real-world hypergraphs

| Hypergraphs | $n$ | $m$ | $|V_{GCC}|$ | $\langle d_N \rangle$ | $\langle d_H \rangle$ | $\langle k_1 \rangle$ | $\langle k_2 \rangle$ |
|---|---|---|---|---|---|---|---|
| Algebra | 423 | 1268 | 420 | 79.45 | 19.66 | 239.07 | 4040.7 |
| Restaurants-Rev | 565 | 601 | 565 | 79.75 | 8.14 | 110.59 | 1105.84 |
| NDC-classes | 1161 | 1088 | 628 | 17.42 | 9.05 | 87.90 | 533.32 |
| iAF1260b | 1668 | 2351 | 1668 | 13.26 | 5.46 | 22.63 | 120.22 |
| iJO1366 | 1805 | 2546 | 1805 | 16.92 | 5.55 | 29.40 | 455.51 |
| Email-W3C | 14317 | 19821 | 13351 | 4.13 | 3.22 | 5.24 | 9.69 |

$\sigma(S)/|V_{GCC}|$ was the average of 100 simulations to ensure the reliability of the results. In these six real-world hypergraphs, the CIA algorithm outperformed other algorithms significantly.

The topological properties of these datasets are shown in Table 1: $n$ and $m$ represent the number of nodes and hyperedges in the hypergraph, $|V_{GCC}|$ is the number of nodes in the GCC, $\langle d_N \rangle$ and $\langle d_H \rangle$ are the average degrees of nodes and hyperedges in the GCC, respectively. $\langle k_1 \rangle$ and $\langle k_2 \rangle$ are the average numbers of 1-simplices and 2-simplices on nodes in the GCC, respectively.

The first case, Algebra, involves interactions between users on a mathematics website. The interactions between users are mainly about comments, questions and answers on algebra problems. Each node represents a user, and users who answer the same type of question (in the area of algebra or geometry) are represented by a hyperedge. From Figure 11(a), it can be seen that the CIA algorithm is significantly better than other algorithms, leading HADP by 2.61% when increased to 40 seeds.

The second case, Restaurant-Rev, has each node and hyperedge representing a user who comments on a particular restaurant and the set of users for that restaurant on the website, respectively. Figure 11(b) shows that HD is the second-best performing algorithm on this dataset, with CIA leading HD by 1.95% at 50 seeds.

In the third dataset, NDC-classes, nodes represent class labels, and hyperedges represent a set of drug class labels. Figure 11(c) shows that HADP is the second-best performing algorithm on this dataset. Similar to Algebra, there are many node pairs that appear in multiple hyperedges, which greatly improves HADP's performance. At 60 seeds, CIA

Influence Maximization in Hypergraphs

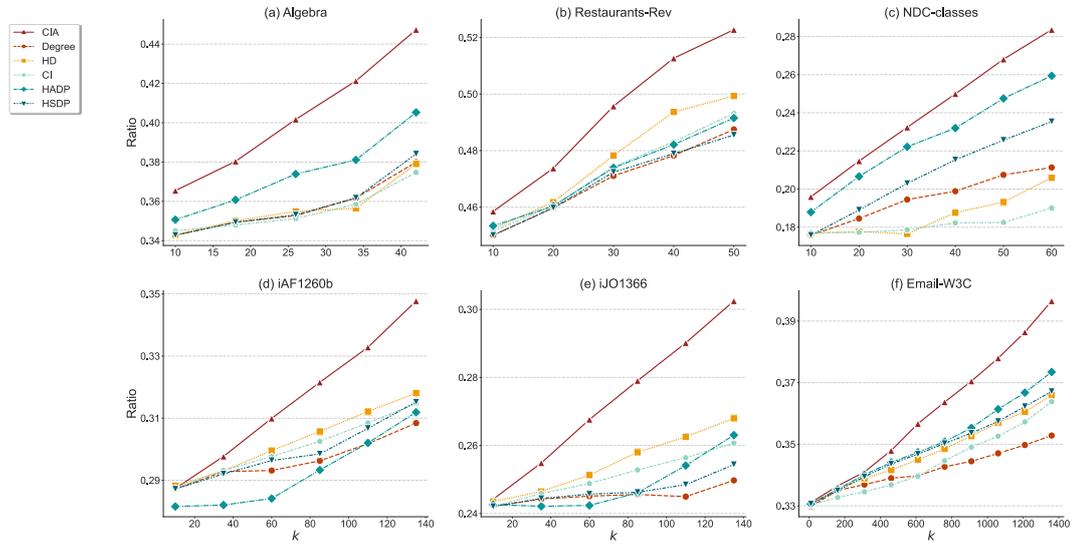

**Figure 11.** In six real-world networks, $\lambda_1 = 1.1$ and $\lambda_2 = 1$, while increasing the number of seeds from 2 (or 10) to 10% of the size of the GCC. These results are the average of 100 independent experiments.

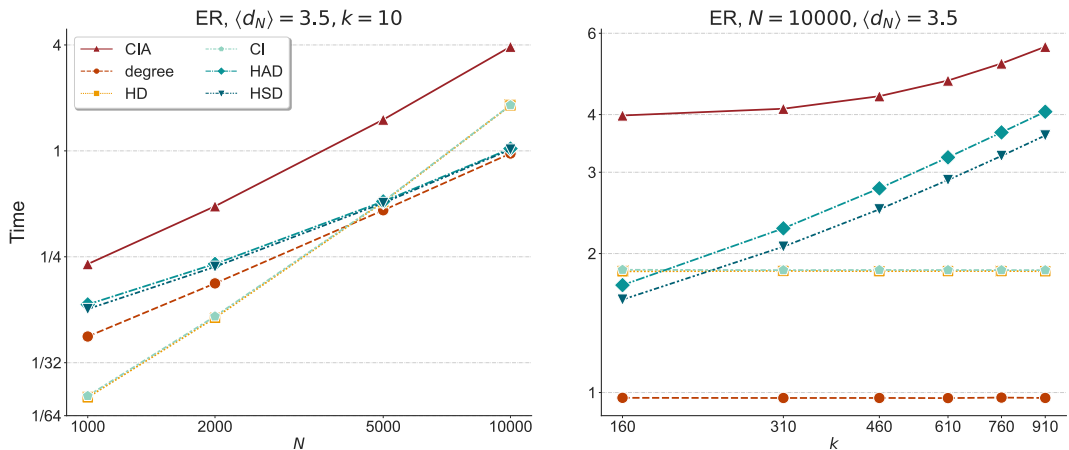

**Figure 12.** The running time of different algorithms on ER hypergraphs with $\langle d_N \rangle = 3.5$. The horizontal axis in the left graph is the network size $N$, while the horizontal axis in the right graph is the seed number $k$. The vertical axis in both graphs is $ln(T)$.

leads HADP by 2.52%.

The fourth and fifth datasets, iAF1260b and iJO1366, have nodes representing reaction-based metabolites and hyperedges representing a set of metabolites applied to a particular reaction. Figure 11 (d) shows that CIA leads HADP by 3.23% at 160 seeds, and Figure 11(e) shows that CIA leads HD by 3.55% at 160 seeds.

In the sixth dataset, Email-W3C, hyperedges are comprised of the sender and all recipients of an email. Figure 11(f) shows that CIA outperforms HADP by 2.31% at 1360 seeds.

Influence Maximization in Hypergraphs

**D.** Computational complexity

We conducted extensive analysis to evaluate the computational complexity of CIA algorithms. Without loss of generality, we ran all considered methods on an ER hypergraph with an average degree $\langle d_N \rangle = 3.5$, and varied the network size $N$ and the number of seeds $k$. Due to more computations, the CIA had the longest running time among all tested methods. We recorded the running time of each algorithm on the ER hypergraphs, and calculated the logarithm ($ln$) of the algorithm running time and the hypergraph size, as shown in Figure 12. We used a first-order polynomial to fit the running time of different algorithms, and found that the time consumed by each algorithm increased linearly with the network size. The time complexity of the HD algorithm was the highest, about $O(N^{1.66})$. The time complexity of the CIA algorithm was about $O(N^{1.24})$, and its running time increased linearly with the network size.

**V. Conclusions**

The Hypergraph Influence Maximization problem has a wide range of application prospects in reality. In this study, we start with a simple propagation model, the Simplicial Contagion Model, and establish the SIR model using the message-passing process. Subsequently, leveraging stability analysis, we articulate the CI metric for node centrality within hypergraphs, forming the basis for the Hypergraph Influence Maximization problem. Our novel contribution involves the development of the CIA algorithm designed for identifying influential spreaders in the propagation dynamics. Experimental evidence on synthetic and real-world networks demonstrates the superiority of the proposed algorithm over commonly used algorithms, validating its effectiveness across various network structures, different parameter combinations, and diverse initial conditions.

However, there are some potential drawbacks to this method. Firstly, the method is applicable to local tree-like networks, and when there are many short cycles in the network, the message-passing equations become less accurate, which may undermine the performance of collective influence in these networks. Secondly, when $\beta_1 = 0$, the dynamical system is non-hyperbolic, which should be incorporated into the framework in future work.

**ACKNOWLEDGMENTS**

This work was supported by National Natural Science Foundation of China (Grant No.12371516), Liaoning Provincial Natural Science Foundation (Grant No.2022-MS-152), Fundamental Research Funds for the Central Universities (DUT22LAB305), National Key Research and Development Program of China (2021ZD0112400, 2020YFA0713702).

**APPENDIX**

**A.** Competing methods

We compared the adaptive algorithm based on collective influence with several widely used metrics for node spread ability ranking.

- Degree-based ranking: In degree ranking, the score of each node $i$ is determined by the number of its neighbors: $d_N(i) = \sum_{j=1}^{N} \tilde{A}_{ij}$.

- Hyperdegree-based ranking: In hyperdegree ranking, the score of each node $i$ is determined by its hyperdegree: $d_H(i) = \sum_{\alpha=1}^{M} I_{i\alpha}$.

- Collective Inference: Extension of collective influence algorithm on hypergraphs by replacing degree with hyperdegree. The score of node $i$'s $CI(i)$ is defined as: $CI(i) = (d_H(i) - 1) \sum_{v_j \in \partial Ball(v_i, l)} (d_H(j) - 1)$, where $Ball(v_i, l)$ is a node set containing all nodes within a radius of $l$, and $l$ represents the shortest path from a node in $Ball(v_i, l)$ to node $v_i$ [36]. The boundary of $Ball(v_i, l)$ is denoted by $\partial Ball(v_i, l)$. In this paper, the parameter $l$ is set to 1.

- Hyper Adaptive Degree Pruning (HADP): Punish the neighbours around the selected seeds to reduce the probability of adjacent nodes being selected as seeds to avoid overlapping influence. In the $k - th$ step, select the node $i$ with the current maximum degree as the seed and update the degree of the nodes: $d_N{}^k(j) = d_N{}^{k-1}(j) - |N_s(i)|$,

Influence Maximization in Hypergraphs

where $N_S(i)$ is the adaptive neighborhood set of the adjacent node $v_j$ of the newly added seed $v_i$ [36].

- Hyper Single Degree Pruning (HSDP): This algorithm considers giving a single punishment to each node during iteration, that is, in the $k-th$ step, $d_N^k(j) = d_N^{k-1}(j) - 1$ [36].

- Random: Randomly select $k$ nodes as seeds in the giant connected component.

**B.** Additional experiments

Figure B1 and Figure B2 respectively show the proportion and distribution of infected nodes in the SF hypergraph. Figure B3 and Figure B4 respectively show the proportion and distribution of infected nodes in the ER hypergraph. Figure B5 shows the performance of different algorithms on $d$-uniform hypergraphs, and the gap between various algorithms is significantly reduced compared to SF hypergraphs and ER hypergraphs. HADP performs the best, followed by CIA.

**C.** Network data

Network data are downloaded from the following websites.
(1) https://www.cs.cornell.edu/~arb/data/.
(2) http://bigg.ucsd.edu/.

Influence Maximization in Hypergraphs

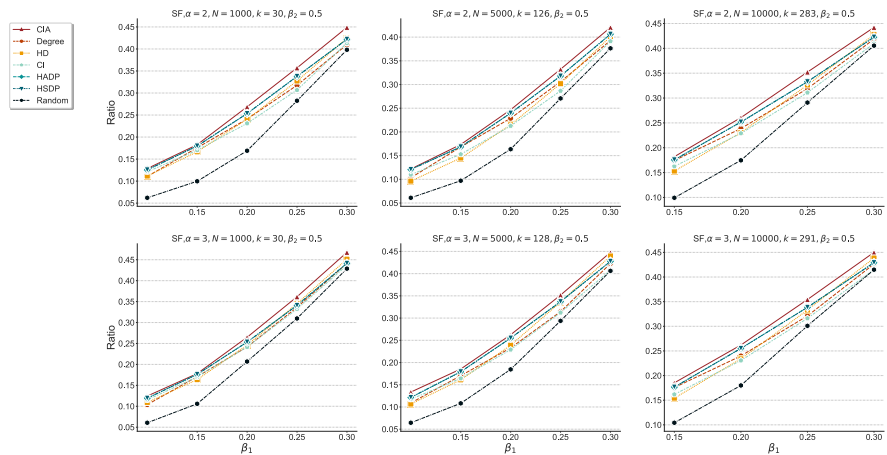

**Figure B1.** The proportion of affected nodes with $\beta_2 = 0.5$ in SF hypergraphs of size $N = 1000, 5000$ or $10000$ and $\alpha = 2$ or $3$. We use different methods (including CIA, degree, HD, CI, HADP, HSDP and Random) to select seeds of 3% of the size of GCC. These results are the average

of 100 independent experiments.

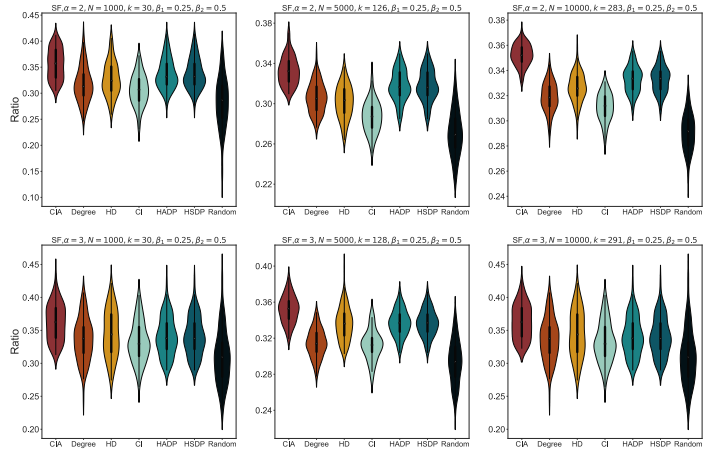

**Figure B2.** The distribution of the proportion of affected nodes with $\beta_1 = 0.3$ and $\beta_2 = 0.5$ in SF networks of size $N = 1000, 5000$ or $10000$ and $\alpha = 2$ or $3$. We use different methods (including CIA, degree, HD, CI, HADP, HSDP and Random) to select seeds of 3% of the size of the GCC. These results are based on 100 independent experiments.

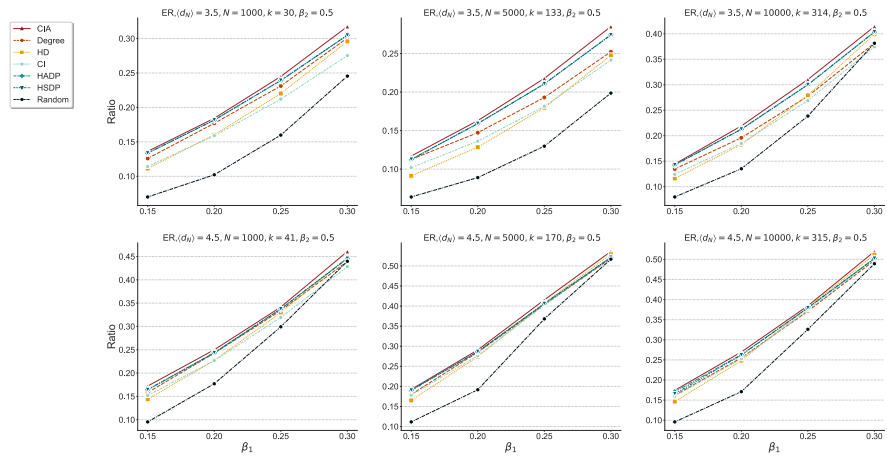

**Figure B3.** The proportion of affected nodes with $\beta_2 = 0.5$ in ER hypergraphs of size $N = 1000, 5000$ or $10000$ and $\langle d_N \rangle = 3.5$ or $4.5$. We use different methods (including CIA, degree, HD, CI, HADP, HSDP and Random) to select seeds of 3% of the size of GCC. These results are the average of 100 independent experiments.

Influence Maximization in Hypergraphs

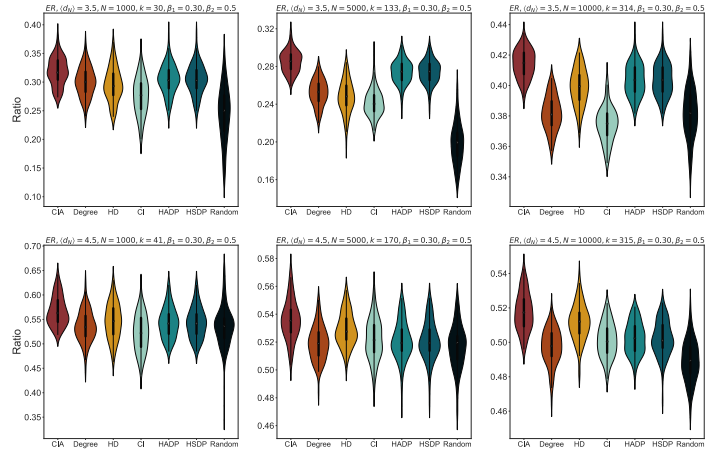

**Figure B4.** The distribution of the proportion of affected nodes with $\beta_1 = 0.30$ and $\beta_2 = 0.5$ in ER networks of size $N = 1000, 5000$ or $10000$ and $\langle d_N \rangle = 3.5$ or $4.5$. We use different methods (including CIA, degree, HD, CI, HADP, HSDP and Random) to select seeds of 3% of the size of GCC. These results are based on 100 independent experiments.

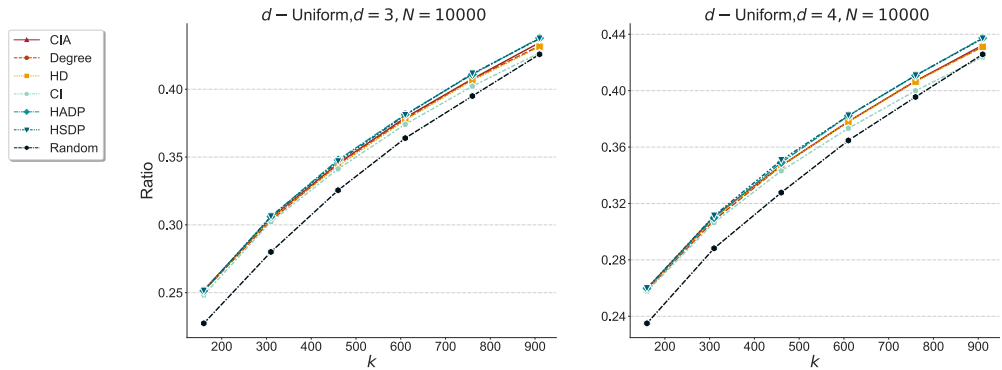

**Figure B5.** The proportion of infected nodes with increasing seed numbers in $d$-Uniform hypergraph of size $N = 10000$ with $d = 3$ or $4$, and fixed $\lambda_1 = 1$ and $\lambda_2 = 1$. These results are the average of 100 independent experiments.